\documentclass[11pt]{article}
\usepackage{moriond,epsfig}

\def\be{\begin{equation}}
\def\ee{\end{equation}}
\def\bea{\begin{eqnarray}}
\def\eea{\end{eqnarray}}

\usepackage{relsize}
\def\babarsym{\mbox{\slshape B\kern-0.1em{\smaller A}\kern-0.1em
    B\kern-0.1em{\smaller A\kern-0.2em R}}}
\begin{document}
\vspace*{4cm}
\title{MEASUREMENTS OF THE CKM ANGLE
$\beta/\phi_1$ AT THE B FACTORIES}

\author{ J. OCARIZ }

\address{
Laboratoire de Physique Nucl\'eaire et des Hautes Energies,
 IN2P3-CNRS- Universit\'es de
Paris VI et VII \\ 4 place Jussieu 75252 Paris cedex 05, France \\
Representing the \babarsym \  and Belle Collaborations}

\maketitle\abstracts{
We report measurements of time-dependent CP asymmetries related to the CKM angle $\beta/\phi_1$,
using decays of neutral B
mesons to charmonium, open charm and in $b\to s$ loop-dominated processes. A preliminary
measurement of time-dependent CP asymmetries in $B^0\to \rho^0(770) K^0_S$ decays from the \babarsym \ 
experiment is given here.
}

\section{Introduction}
In the Standard Model (SM), CP violation occurs as a consequence of a complex phase in the 
$3\times 3$ Cabibbo-Kobayashi-Maskawa (CKM) mixing matrix~\cite{ref:CKM}. 
The unitarity of the CKM matrix imposes the condition $V_{ub}^*V_{ud}+V_{cb}^*V_{cd}+V_{tb}^*V_{td}=0$,
where $V_{ij}$ are the CKM matrix elements.  
This condition can be conveniently illustrated as a triangular relation in the
$(\bar{\rho},\bar{\eta})$ complex plane. A non-vanishing phase in the
CKM matrix results in a non-zero area for the Unitarity Triangle (UT). 
Various measurements in the $B$ meson
system are sensitive to the CKM angles
$\alpha$, $\beta$, and $\gamma$ of the 
UT~\footnote{ 
The CKM angles are also called $\phi_2$, $\phi_1$ $\phi_3$.
In this report we will indistinctly use either convention.}.

In the $B^0-\bar{B}^0$ system, information on the CKM  angles
can be obtained by measuring the time dependence of  $B^0$ or $\bar{B}^0$ decays
 to CP eigenstates $f_{CP}$. The time distribution is given by
\begin{equation}
\frac{dN(B^0 (\bar{B}^0)\to f_{CP})}{dt} \propto
e^{- t/\tau}
\left[
	1 -
	\left(
		\pm S_f\sin{\Delta mt} \mp C_f\cos{\Delta mt}
	\right)
\right],
\end{equation} 
where $\tau$ is the $B^0$ meson
lifetime, and $\Delta m$ is the
$B^0-\bar{B}^0$ oscillation frequency. The CP-violating coefficients 
$S_f$ and $C_f$ are functions of the parameter
$\lambda_f$: 
\begin{equation}
	\lambda_f  = \frac{q}{p}\frac{A(\bar{B}^0\to f)}{A(B^0\to f)}  \ , \ 
        S_f=\frac{2Im(\lambda_f)}{1+|\lambda_f|^2}  \ , \
         C_f=\frac{1-|\lambda_f|^2}{1+|\lambda_f|^2}.
\end{equation}
In this expression $A(B^0\to f)$ (resp. $A(\bar{B}^0\to f)$) is the decay amplitude of $B^0$ 
(resp. $\bar{B}^0$) to the 
final state $f_{CP}$ respectively, and the $q/p$ ratio
ratio is given by  the admixture of flavour 
eigenstates $B^0$ and $\bar{B}^0$ in the neutral $B$ mass eigenstates.  The SM predicts $|q/p|\simeq 1$.
If only one weak phase enters
the decay amplitude,  $\lambda_f=\eta_f e^{2\theta}$,
where $\eta_f = \pm 1$ is the CP of the final state $f$. 

The CP-violating parameter
$\sin{2\beta}$ is most accurately measured using $B^0\to J/\psi K^0$ decays. These decays are dominated 
by a $b\to c\bar{c}s$ tree amplitude and a $b\to sc\bar{c}$ penguin amplitude. As 
contributions with different weak phases
are doubly Cabibbo-suppressed, the CP violation parameters are
$S=\pm\sin{2\beta}$ and $C=0$ to an excellent approximation. 
Other measurements related to the CKM angle $\beta$ are given by the 
$b\to s$ transition decays.
In the SM, these decays occur dominantly through pure penguin diagrams, and the CP phase originally acquired
in the $B^0-\bar{B}^0$  mixing is not changed. If new particles contribute
in the loop, they introduce new couplings, and the corresponding new phases 
will shift the CP asymmetry parameter $\sin{2\beta_{eff}}$ from its SM value $\sin{2\beta}$. 
Therefore the measurement of $\sin{2\beta_{eff}}$ for such decays are a potentially
sensitive probe to New Physics.
\begin{figure}
\psfig{figure=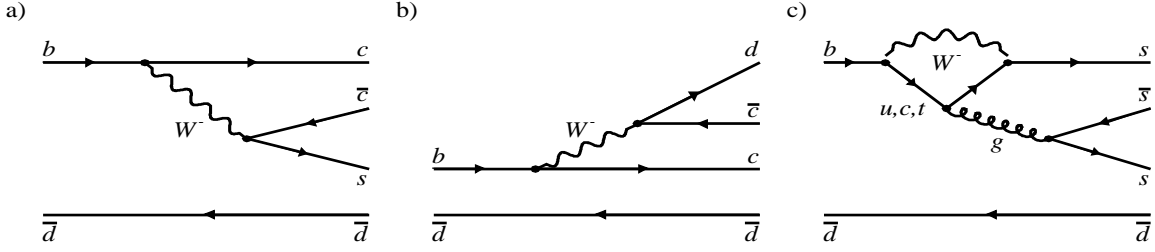,width=0.95\linewidth,height=0.20\linewidth}
\caption{Diagrams mediating the decay amplitudes related to the measurements of the CKM angle
$\beta$. From left to right: a) $b\to c\bar{c}s$ (``charmonium''), b) $b\to c\bar{c}d$ (``open charm'')
and c) $b\to q\bar{q}s$ (``penguin-dominated'').
\label{fig:feynman}}
\end{figure}
\section{Experimental Technique}
The results presented are from the \babarsym \ and Belle experiments. \babarsym \ runs at the 
PEP-II asymmetric energy $e^+e^-$ collider, and the Belle detector is located at the KEKB 
asymmetric collider. 
Both colliders, usually referred to as $B$ factories,  operate at 
the $\Upsilon(4S)$ resonance, whose mass is slightly above the
$B-\bar{B}$ production threshold.
At the B factories, the  center-of-mass reference frame is boosted with respect to the 
detector frame, in the direction of the beam line axis $z$. The boost parameter 
$\beta\gamma$ is 0.55 for PEP-II and 0.425 for KEKB.
The $B-\bar{B}$ pairs being produced almost at rest in the  $\Upsilon(4S)$  rest frame, $\Delta t$ can be
determined from the displacement $\Delta z$ between the $B$ decay vertices:
$\gamma\beta c\Delta t \simeq (z_{CP}-z_{TAG})$, where $z_{CP}$ refers to the vertex of the
$B$ meson fully reconstructed in the final state $f_{CP}$ , and $z_{TAG}$ to the
vertex of the other $B$ meson in the event, whose decay products are also used to identify its flavour
at decay time.

At the B factories, CP violation is studied through the measurement of the time-dependent CP asymmetry, 
$A_{CP}(\Delta t)$. This quantity is defined as
\begin{equation}
A_{CP}(\Delta t) = \frac{N(\bar{B}^0\to f_{CP})-N(B^0\to f_{CP})}{N(\bar{B}^0\to f_{CP})+N(B^0\to f_{CP})}(\Delta t),
\end{equation}
where $N(\bar{B}^0\to f_{CP})$ is the number of $\bar{B}^0$ that decay into the final state $f_{CP}$ 
at the time difference $\Delta t$. In general, this asymmetry can be expressed as the sum of two components:
\begin{equation}
A_{CP}(\Delta t) =  S_f\sin{\Delta m \Delta t} - C_f\sin{\Delta m \Delta t}.
\end{equation}
When only one amplitude contributes to the final state, the cosine term vanishes, as direct CP
violation requires at least two different weak and strong phases to occur. This holds
in particular for decays such
as $B^0\to J/\psi K^0$, where one has
$S_f = -\eta_f\sin{2\beta}$, $\eta_f$ being the CP value of the final state considered
(i.e. $\eta=-1$ for $J/\psi K^0_S$, and $\eta=+1$ for $J/\psi K^0_L$).

Constraints on the CKM angle $\beta$ can be obtained
 through the three types of $B^0$ decays illustrated in
Figure \ref{fig:feynman}: the charmonium modes $b\to c\bar{c}s$, the ``open-charm''modes
$b\to c\bar{c}d $ and the penguin-dominated $b\to q\bar{q}s$ modes. 
They are described in the following sections.
\begin{figure}
\psfig{figure=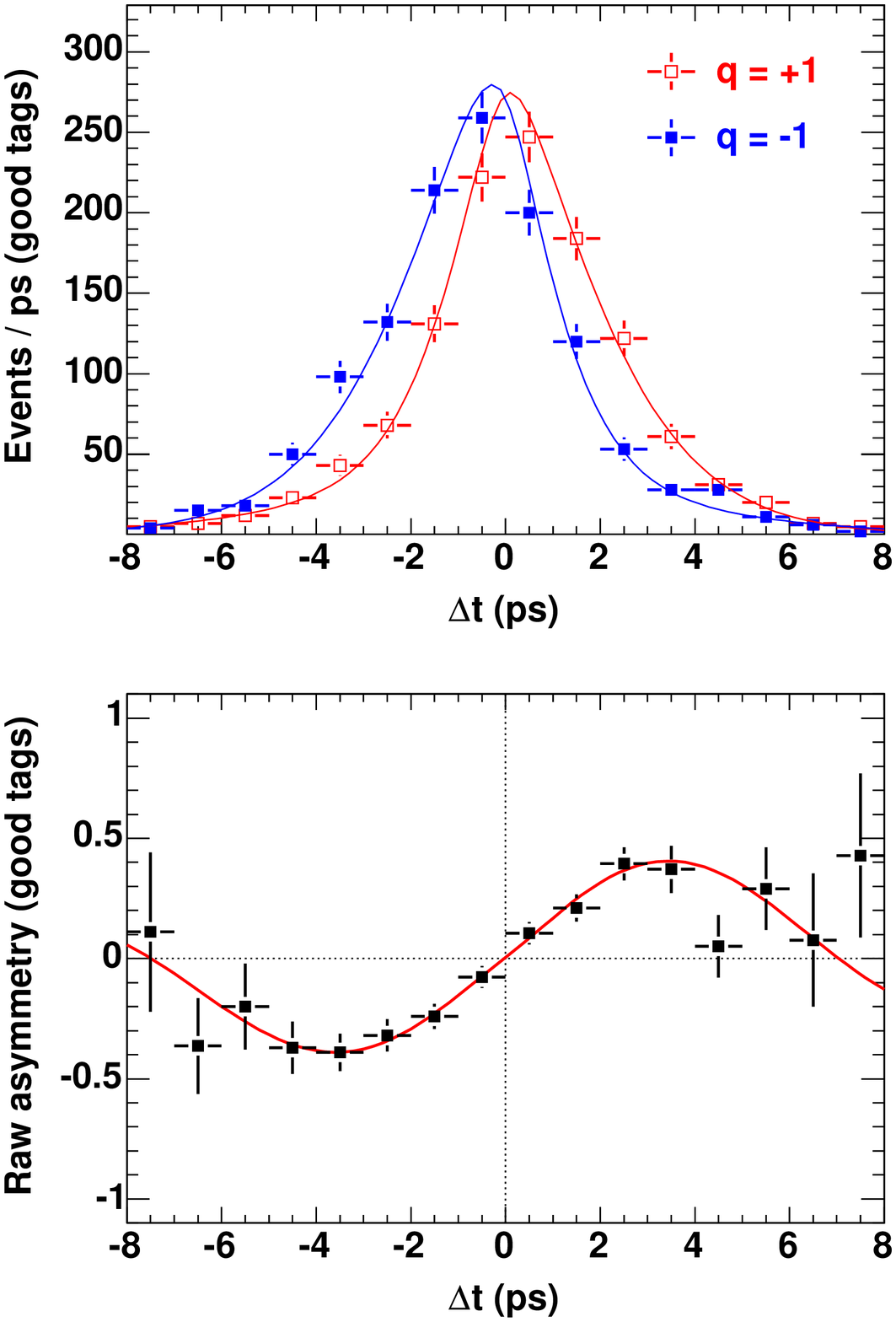,width=0.3\linewidth,height=0.32\linewidth}
\psfig{figure=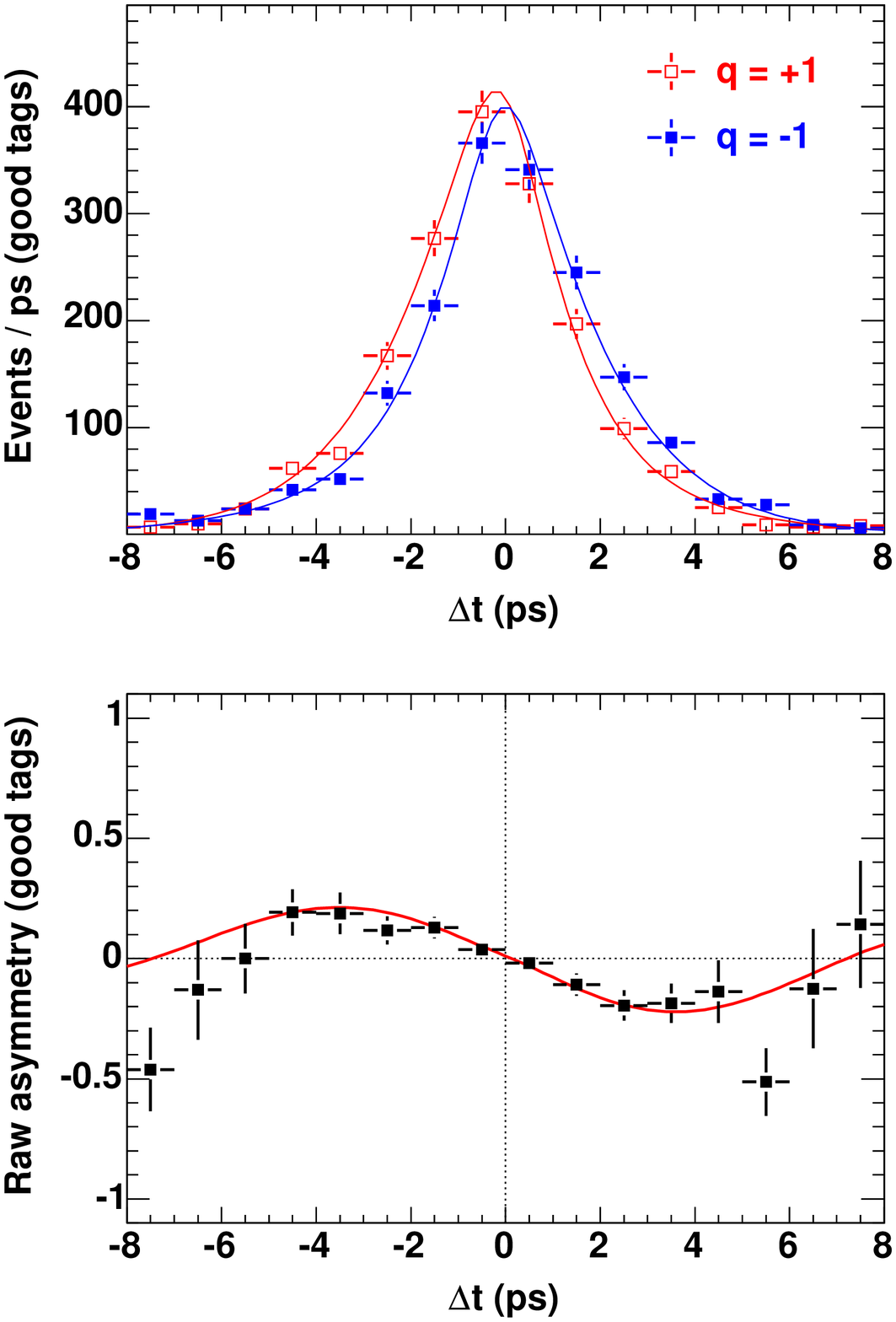,width=0.3\linewidth,height=0.32\linewidth}
\psfig{figure=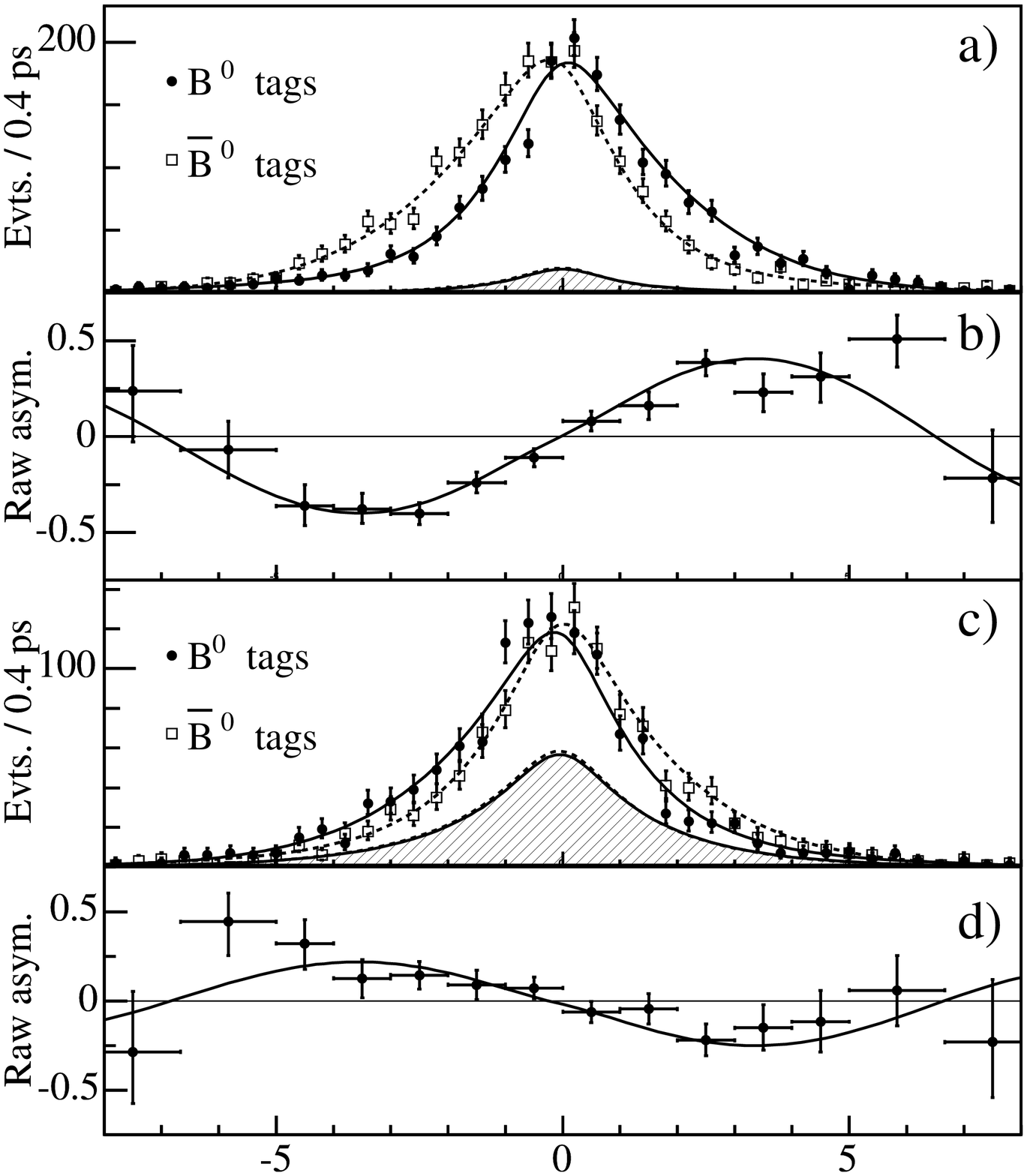,width=0.32\linewidth,height=0.3\linewidth}
\caption{Measurement of $\sin{2\beta}$  in the $b\to c\bar{c}s$ modes, form
 the Belle (left, center) and \babarsym \ (right) experiments. 
The time distributions for events tagged as $B^0$ or $\bar{B}^0$ in CP-odd  or CP-even
final states are shown, together with the corresponding raw CP asymmetry, with the
projection of the unbinned maximum likelihood fit superimposed.
\label{fig:ccbars}}
\end{figure}
\section{The CKM angle $\beta/\phi_1$ from $b\to c\bar{c}s$}
These modes, also known as ``golden'' modes, are dominated by a tree level diagram $b\to c\bar{c}s $.
Furthermore, the leading penguin contribution to the final state has the same weak phase
as the tree, and the largest term with a different weak phase is a doubly Cabibbo-suppressed penguin. 
This makes $C=0$ a very good approximation.

The CP eigenstates considered for this analysis are the $J/\psi K^0_S$, $\psi(2S)K^0_S$,
$\chi_{c1}K^0_S$, $\eta_c K^0_S$ and $J/\psi K^0_L$.
These modes also offer experimental advantages, because of their
relatively large branching fractions allowing to collect large signal samples, 
and the presence of narrow charmonium resonances in the final
states, that provide a clear experimental signature and a strong 
rejection of combinatorial background.

\begin{figure}
\psfig{figure=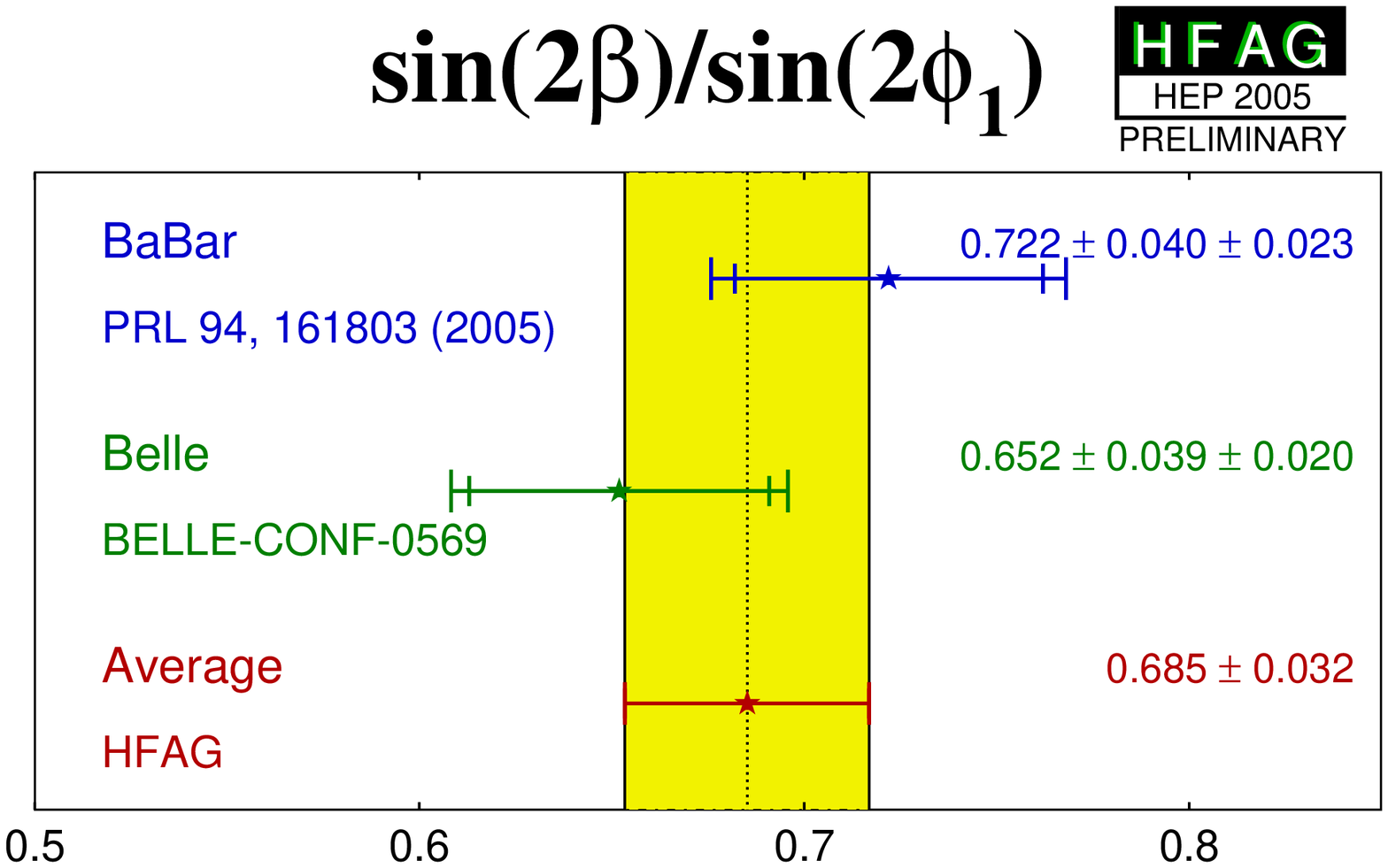,width=0.32\linewidth,height=0.28\linewidth}
\psfig{figure=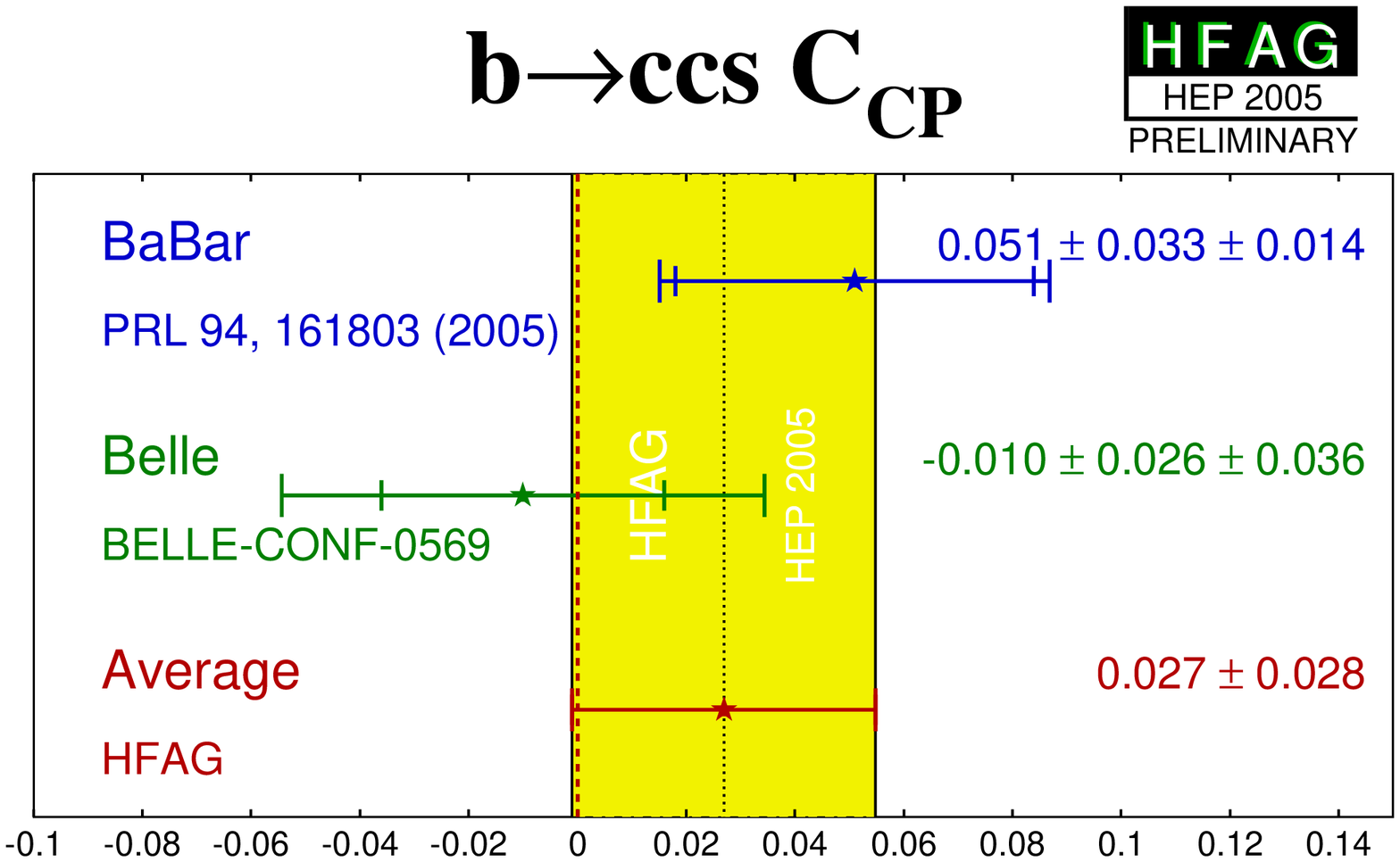,width=0.32\linewidth,height=0.28\linewidth}
\psfig{figure=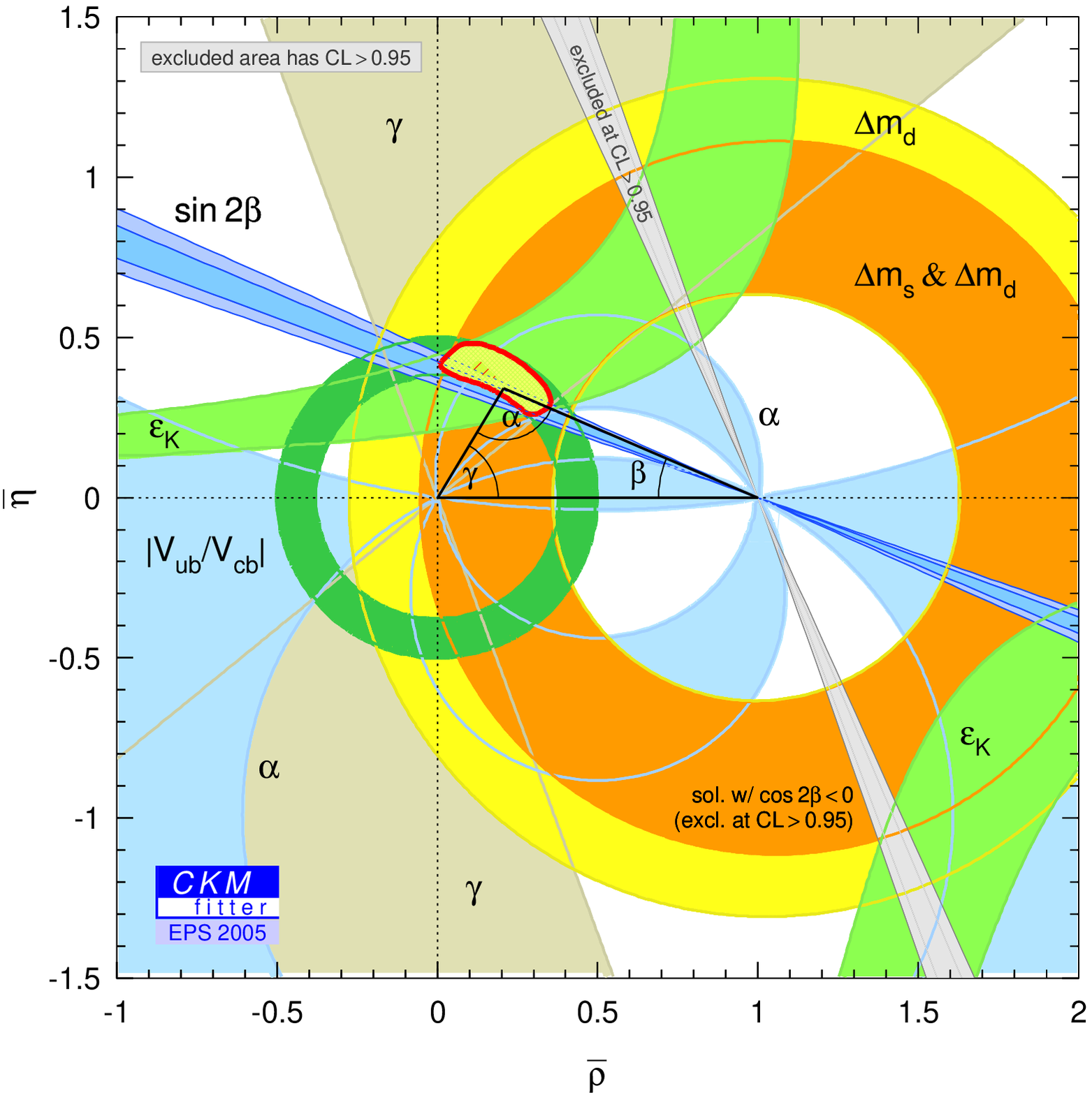,width=0.30\linewidth,height=0.32\linewidth}
\caption{ Average of measurements 
$b\to c\bar{c}s$ processes 
of the CP violating parameters $S$ (left) and $C$ (center), from the \babarsym \ and Belle experiments.
Right:  constraints on the UT, coming from a global fit to the CKM
matrix.
The red curve surrounds the $>95\%$ CL area, from all measurements excluding the direct
measurements of $\sin{2\beta}$, and the light blue contours represent the direct
constraints on the CKM angle $\beta$.
\label{tab:ccbars}}
\end{figure}

The asymmetry between the $\Delta t$ distributions for $B^0$ and $\bar{B}^0$ tagged events,
clearly visible in Figure \ref{fig:ccbars}, 
establishes CP violation in the $B$ meson system.
The average of results are given in the Tables~\cite{ref:hfag} displayed in Figure \ref{tab:ccbars}.
The latest \babarsym \ measurements of $\sin{2\beta}$ are performed on a sample of $227\times 10^6$
$B-\bar{B}$ pairs, and the Belle results use a data sample made of $386\times 10^6$
$B-\bar{B}$ pairs~\cite{ref:sin2beta}.
The world average value for $\sin{2\beta}$, heavily dominated by the results from \babarsym \ and Belle,
is $\sin{2\beta} = (0.685\pm 0.032)$. 
The main sources of systematical errors
are uncertainties in the background level and characteristics, in the parametrisation of time
resolution, and in the measurement of the wrong tagging rates. Most of these systematics
are of statistical nature and
will thus decrease with additional statistics.

The direct measurement of $\sin{2\beta}$ can be compared with the indirect constraints
on the UT~\cite{ref:CKMfitter} originating from World Average values 
of $\varepsilon_K$,$|V_{ub}/V_{cb}|$, $B_d$ and $B_s$
mixing, as illustrated in Figure \ref{tab:ccbars}. 
The excellent agreement between direct measurements of $\sin{2\beta}$ and indirect constraints,
is a strong indication that the CKM mechanism is the dominant source for CP violation.
\subsection{From $\sin{2\beta}$ to $\beta$}
Measurements of the CP violating parameter $\sin{2\beta}$ leave a four-fold ambiguity
in the CKM angle $\beta$ itself. The ambiguities can be reduced 
by measuring the sign of $\cos{2\beta}$. This measurement provides a direct test of the SM, 
since $\cos{2\beta}$ is predicted to be positive. The interference of CP-even and CP-odd
components in the time-dependent angular distribution $B^0\to J/\psi K^{*0}$ decays, with
$K^{*0}\to K^0_S\pi^0$ provides a measurement of $|\cos{2\beta}|$. 
The \babarsym \ experiment has performed an  analysis on a data sample of  
approximately 88 million $B\bar{B}$ pairs,
and measures $|\cos{2\beta}|=(+2.72^{+0.50}_{-0.79}\pm 0.27)$,
with the value of $\sin{2\beta}$  fixed to its measured value~\cite{ref:marc}.  For the Belle experiment, 
275 million $B\bar{B}$ pairs have been used, providing a result of $|\cos{2\phi_1}|=(+0.56\pm 0.79\pm 0.11)$.
Furthermore, the analysis
of the $K^0_S\pi^0$  phase motion can be used as extra input  
to solve the residual ambiguity. 
From comparison with the 
$K\pi$ scattering 
data from the LASS experiments,
only the positive solution for $\cos{2\beta}$ shows  good
agreement with the scattering data,
and a negative value for $\cos{2\beta}$ is excluded at $86\%$  CL.

Other methods to break the ambiguities have been explored.
In particular, it  has been shown~\cite{ref:Bondaretal} that 
in $B^0 \to D h^0$ with multi-body $D$ decays ($b\to c\bar{u}d$
processes), a time-dependent analysis of the Dalitz plot of the $D$ decay 
allows a direct determination of 
the CKM angle $\phi_1$. The result from the Belle experiment is $\phi_1=(16\pm 21\pm 12)^o$,
and this rules out the $\phi_1=68^o$ solution at $97\%$ CL~\cite{ref:bellebcud}.
\section{Open-charm decays }
The decays $B^0\to D^{(*)+}D^{(*)-}$ are dominated by a tree level diagram $b\to c\bar{c}d$.
Also belonging to the $b\to c\bar{c}d $ class is the $B^0\to J/\psi \pi^0$ mode~\cite{ref:jpsipi0}.
In the SM, the leading penguin contribution to the latter decay
is expected to be small. Corrections for the penguin-induced shifts to $S=\sin{2\beta}$
and $C=0$
have been estimated 
to be a few percent. New Physics beyond the SM could enlarge the penguin 
contribution and would lead to a measurement of 
time-dependent CP asymmetries significantly different from 
the one measured in $b\to c\bar{c}s$ modes. Probing the tree-dominance scenario is an
interesting test of the SM. 

\begin{figure}
\psfig{figure=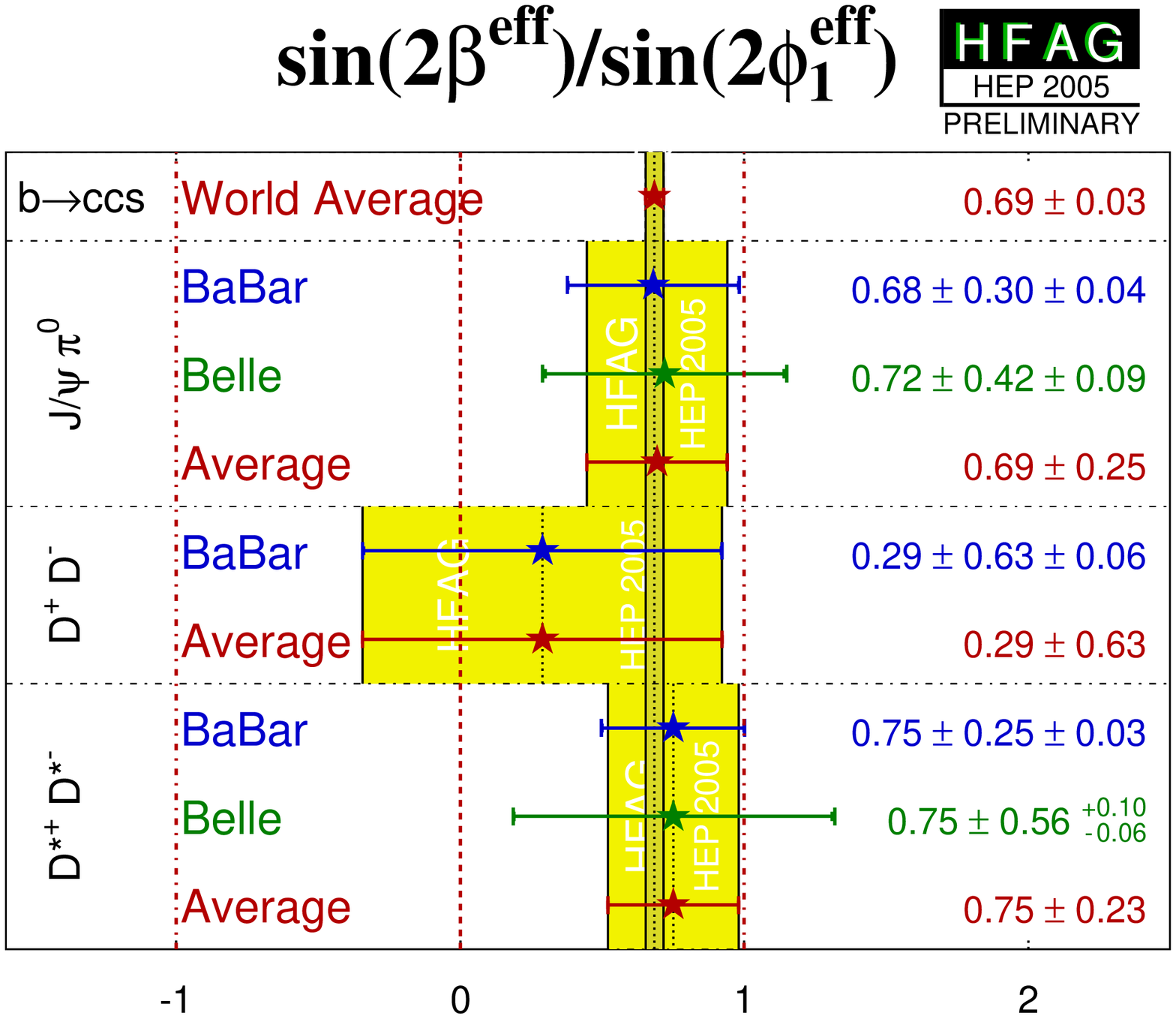,width=0.41\linewidth,height=0.30\linewidth}
\psfig{figure=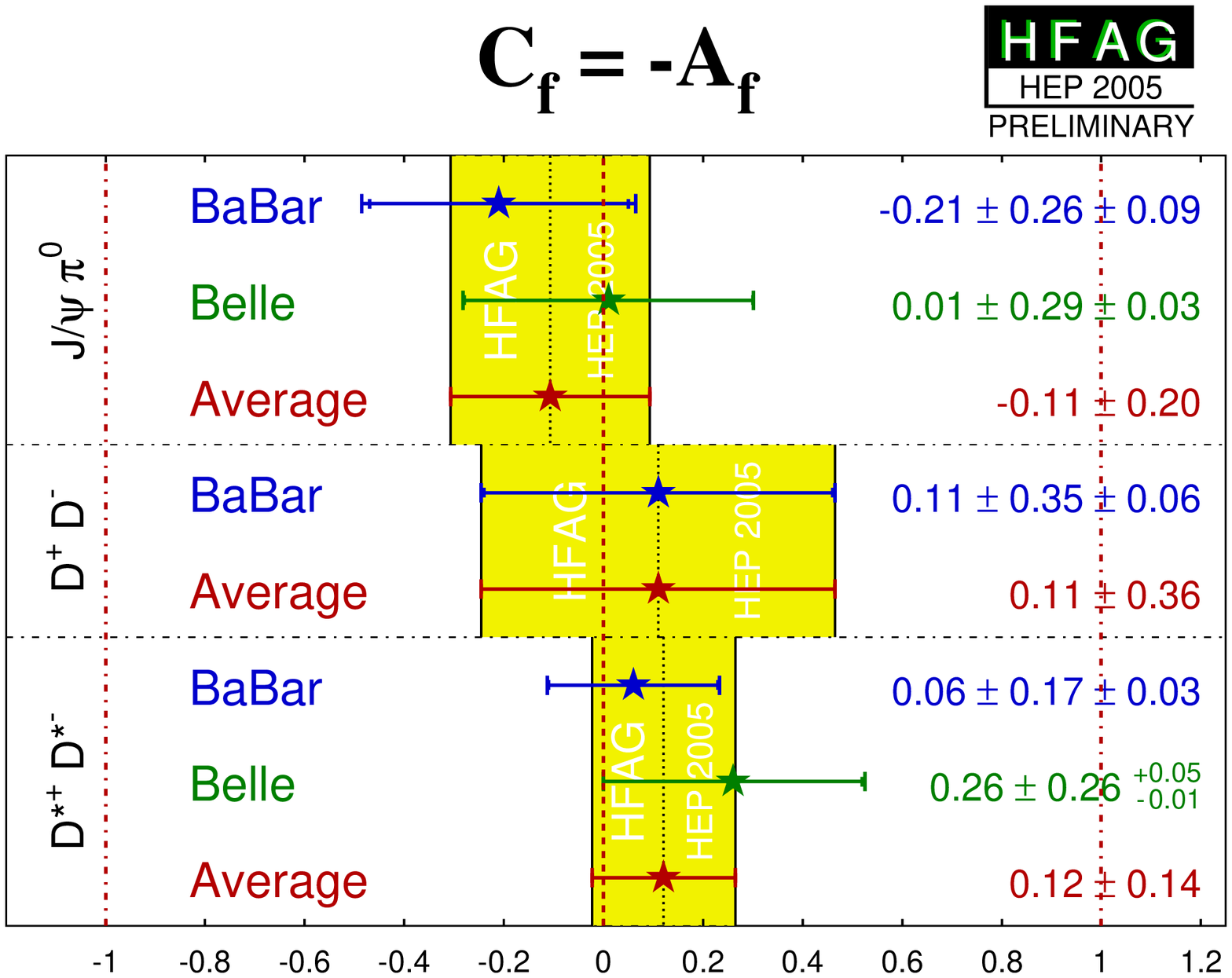,width=0.44\linewidth,height=0.30\linewidth}
\caption{
Summary of results on CP violation for the $b\to c\bar{c}d$ processes. All results on 
$S$ are compatible with the value of $\sin{2\beta}$ from $b\to c\bar{c}s$ processes,
and no evidence for direct CP violation is observed in the $C$ measurements. 
\label{fig:ccbard}}
\end{figure}

The extraction of the coefficients $S$ and $C$ is straigthforward in the $D^+D^-$ and $J/\psi \pi^0$ modes,
which are pure CP eigenstates. In constrast, as the $D^*D^*$ modes are an admixture of CP-odd and CP-even
components,  the CP-odd fractions need to be measured for the extraction of $S$ and $C$ parameters. 
The \babarsym \ experiment has performed a transversity analysis giving a CP-odd fraction
$R_T = ( 0.125 \pm 0.044\pm 0.007)$, and the corresponding result for the Belle experiment~\cite{ref:bdd}
is $R_T=(0.19\pm 0.08\pm 0.01)$.
The results of the CP fits for  all $b\to c\bar{c}d$
modes are summarised in the Table~\cite{ref:hfag} represented in Figure~\ref{fig:ccbard}. Results are all 
consistent with the tree-dominance scenario;
while small deviations  are
expected, they  still lie below experimental sensitivity, and more data is required for 
testing the SM.
\section{Penguin-dominated decays}
In the SM, final states dominated by a $b\to s\bar{s}s$ or a $b\to d\bar{d}s$  transition
offer an independent test of the SM, by comparing the CP-violating parameters in loop processes
with those obtained in the tree-dominated ones. Examples of final states related to the
$b\to q\bar{q}s$ processes are the 
$\phi K^0$, $\eta^\prime K^0$, $f_0K^0$, $\pi^0K^0$, $\omega K^0$, $K^+K^-K^0$ 
and $K^0K^0K^0$ modes~\cite{ref:qqbars}.
These decays are dominated by gluonic penguin amplitudes; depending on the modes considered,
other contributions to the amplitudes may  be present.
In presence of non-SM physics, new particles
could contribute to the loop amplitudes and afffect the time-dependent asymmetries.

\begin{figure}
\psfig{figure=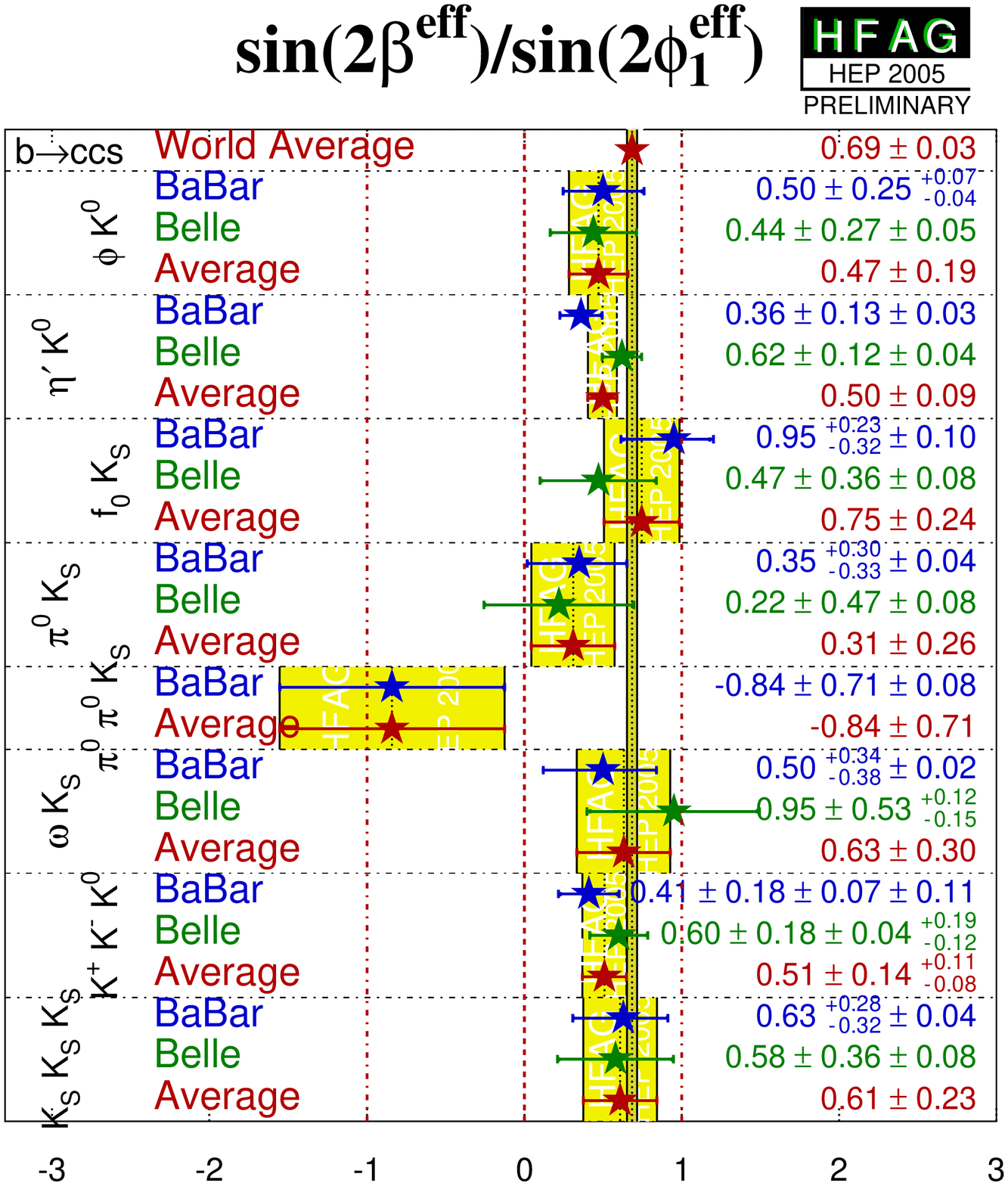,width=0.34\linewidth,height=0.29\linewidth}
\psfig{figure=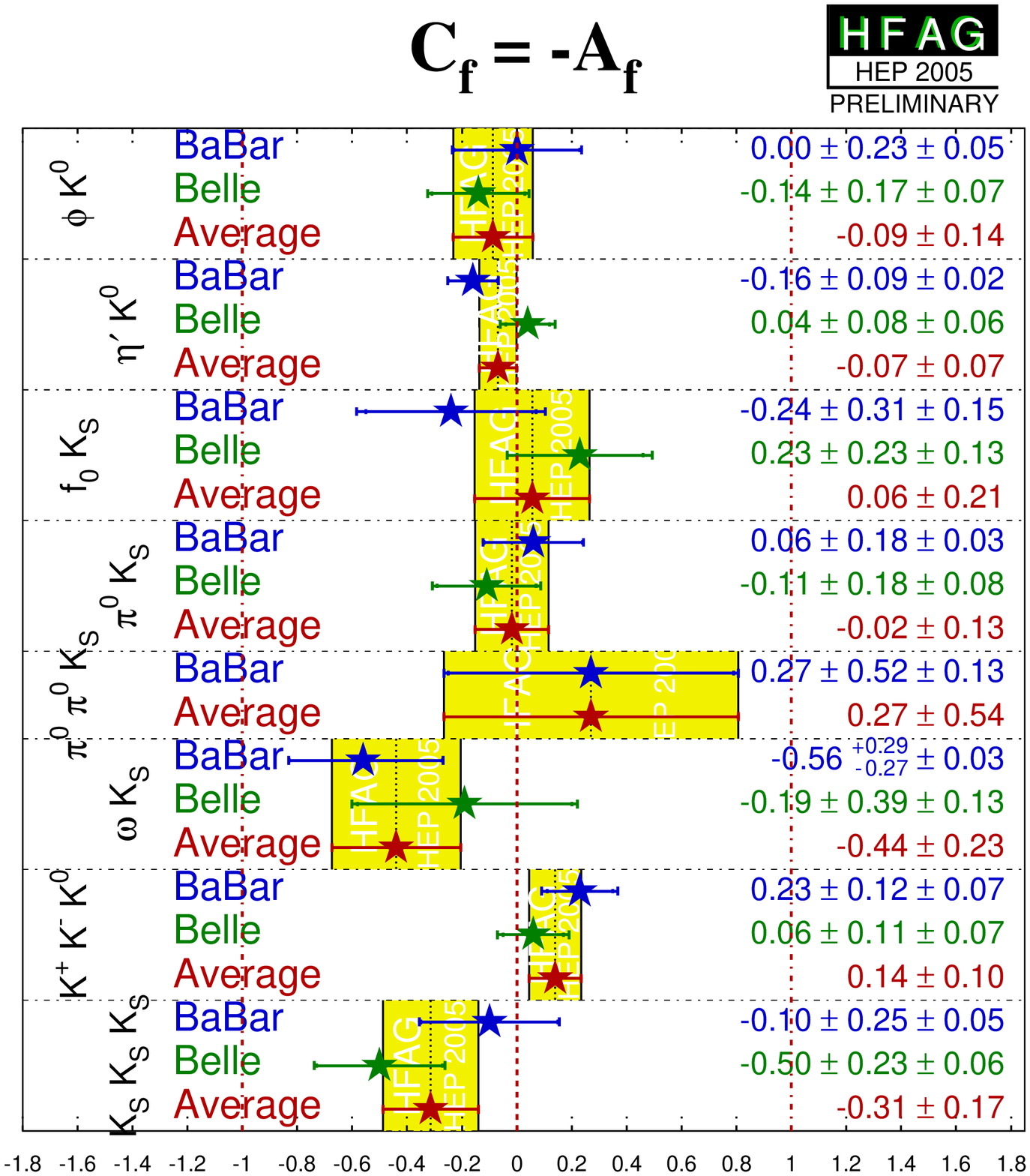,width=0.31\linewidth,height=0.29\linewidth}
\psfig{figure=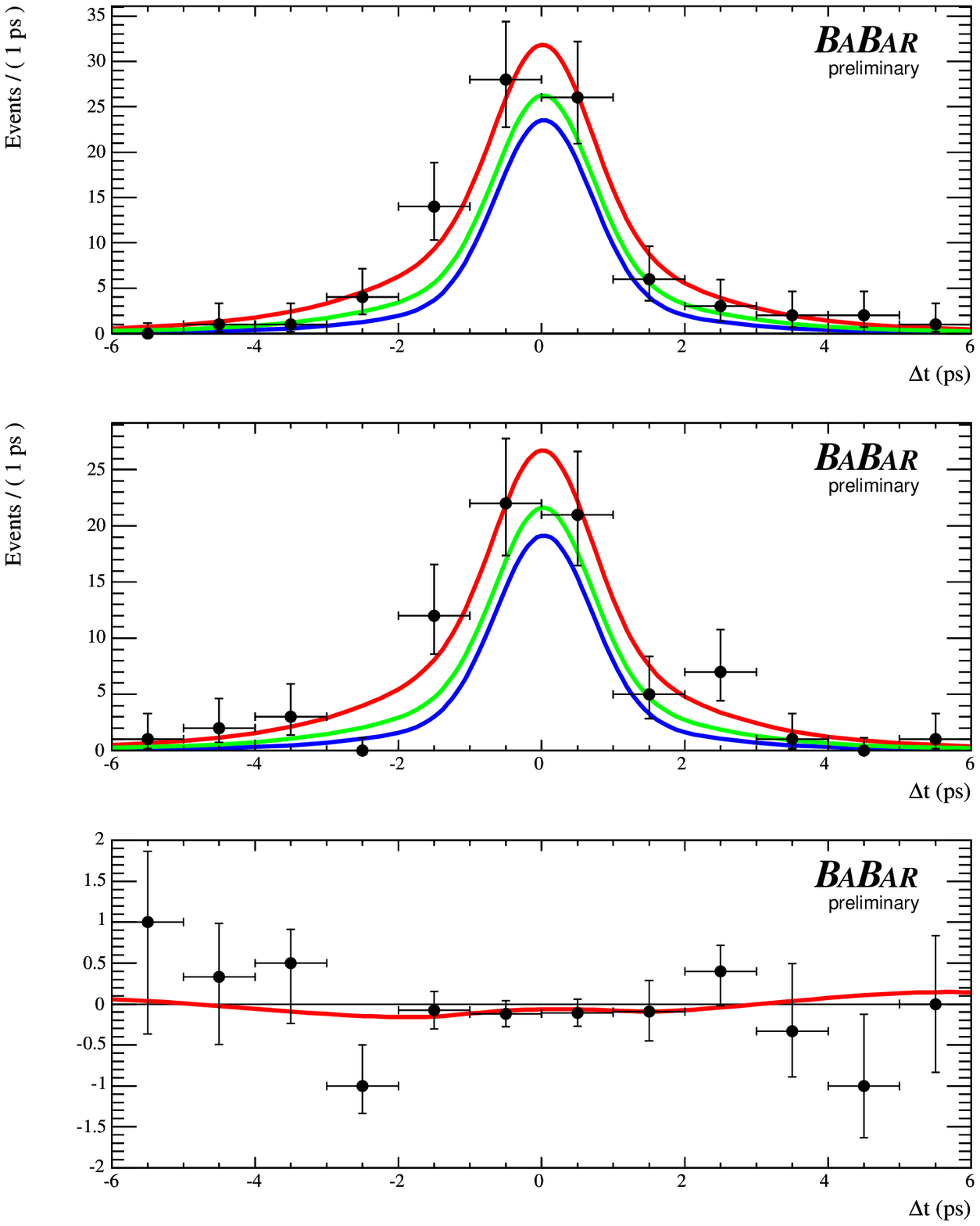,width=0.33\linewidth,height=0.29\linewidth}
\caption{ Averages of measurements of the CP-violating parameters $S$ (left) and $C$ (center) in the
$b\to q\bar{q}s$ modes. No significant deviation from the $S=\sin{2\beta}$, $C=0$ values 
from $b\to c\bar{c}s $ is observed.
Right: (preliminary result from the \babarsym \ experiment) time distributions for $\rho^0(770)K^0_S$ events, tagged as $B^0$ (top), or
$\bar{B}^0$ (center), and raw CP asymmetry (bottom). 
\label{fig:qqbars}}
\end{figure}

The decays $\phi K^0$ provide an example of these studies. In the SM, these 
decays are almost pure
$b\to s\bar{s}s$  penguins, and their CP asymmetry is expected to coincide with the one measured in $b\to c\bar{c}s$
within a few percent. Experimentally, this cannel is also very clean, owing to the narrow
$\phi$ resonance. Although the experimental signature is less clean, the $B^0\to \phi K^0_L$ final
state is also studied.
The decay $B^0\to \eta^\prime K^0$, with its relatively large branching fraction
($\sim 6\times 10^{-5}$) has drawn attention as well. In the SM, the $\eta^\prime K^0$ decay is
expected to be dominated by penguin-mediated amplitudes, and deviations from $S=\sin{2\beta}, C=0$
are therefore expected to be small. Experimental measurements include both the $\eta^\prime K^0_S$ and
$\eta^\prime K^0_L$ modes, these last providing an additional 50\% data sample.
More challenging, the $B^0\to \pi^0K^0_S$ and $B^0\to K^0_SK^0_SK^0_S$ modes
are studied by means of a beam-spot constraint
technique for the measurement of the $B^0$ vertex.  This technique is validated with a control
sample of $B^0\to J/\psi K^0_S$  decays, ignoring the tracks from the $J/\psi$ in the vertexing.
The measurement of the $B^0$ lifetime provides a supplementary validation.
The \babarsym \ experiment has also performed a preliminary measurement of time-dependent CP asymmetries
in the $B^0\to \rho^0(770)K^0_S$ mode. While there is a tree contribution to the amplitude,
it is both colour- and Cabibbo-suppressed, and deviations from $S=\sin{2\beta}$ are expected to be small.
The $\Delta t$ distributions for events tagged as $B^0$ or $\bar{B}^0$, and the CP asymmetry $A_{CP}$,
are shown in Figure~\ref{fig:qqbars}.
The preliminary values are $S=(0.17\pm 0.52\pm 0.26)$, and $C=(0.64\pm 0.41\pm 0.25)$.

The Table~\cite{ref:hfag} in Figure~\ref{fig:qqbars}  summarises the measurements 
of $b\to s$-penguin modes available (the $\rho^0(770)K^0_S$
mode not being included in the Table). While no significant discrepancies with respect to the 
$S=\sin{2\beta}$, $C=0$ values are observed in a mode-by-mode basis, a naive averaging of the $S$ and $C$
parameters should be interpreted with extreme caution; each mode is in principle sensitive to
different SM contributions, so shifts in $\Delta S=\sin{2\beta_{eff}}-\sin{2\beta}$ and
$\Delta C$ are in principle different for each mode. 
More data is therefore required for drawing conclusions. With increasing data samples
integrated by the B factories, new modes from the
$b\to q\bar{q}s$ class become accessible, and provide further tests of the SM.
 \section*{Conclusions}
 The CP-violating parameter 
 $\sin{2\beta}$ in charmonium decays 
 is now measured to a $5\%$ accuracy.
 The measured value is consistent
 with SM expectations. No deviation from the ``tree-dominance'' scenario has been observed in  open-charm decays.
 Time-dependent CP asymmetries in loop-dominated processes are potentially sensitive to
 contributions from New Physics beyond the SM. Measurements being statistically limited, further studies
 with larger datasets are required. 

\end{document}